%% file: ms.tex
\documentclass[a4paper,12pt]{article}%

\title{Technical Report - Milestone 1.1}
\author{Prof. Hans P. Reiser, Johannes K\"ostler}
\date{\today}

\usepackage[left=1in,right=1in]{geometry}
\usepackage[T1]{fontenc}
\usepackage[utf8]{inputenc}
\usepackage[english]{babel}
\usepackage{amsmath}%
\usepackage{amsfonts}%
\usepackage{amssymb}%
\usepackage{titlesec}

\usepackage{enumitem}

\normalsize

\usepackage[numbers,square,sort]{natbib}

\usepackage{graphicx}
\usepackage{svg}

\usepackage{color}
\definecolor{lightgray}{gray}{0.95}
\definecolor{commentgray}{gray}{0.75}
\definecolor{darkgray}{gray}{0.3}
\definecolor{purple}{rgb}{0.65, 0.12, 0.82}

\usepackage{lastpage}
\usepackage[hyphens]{url}
\usepackage[]{hyperref}
\hypersetup{colorlinks = {true}, citecolor = {black}, filecolor = {black},
linkcolor = {black}, menucolor = {black}, urlcolor = {black}}

\usepackage{longtable}
\usepackage{booktabs}
\usepackage{makecell}

\usepackage{listings}
\usepackage{txfonts}
\lstdefinelanguage{JavaScript}{
 keywords={typeof, new, true, false, catch, function, return, null, catch, switch, var, if, in, while, do, else, case, break},
 keywordstyle=\color{black}\bfseries,
 ndkeywords={class, export, boolean, throw, implements, import, this},
 ndkeywordstyle=\color{darkgray}\bfseries,
 identifierstyle=\color{black},
 sensitive=false,
 comment=[l]{//},
 morecomment=[s]{},
 commentstyle=\color{commentgray}\ttfamily,
 stringstyle=\color{black}\ttfamily,
 morestring=[b]',
 morestring=[b]"
}
\usepackage{expdlist}
\usepackage[justification=raggedright,hypcap=false,singlelinecheck=true,
format=hang,labelfont=bf,textfont=normal]{caption}

\usepackage{todonotes}

\usepackage{geometry}
\makeatletter
\renewcommand{\maketitle}{
\begin{titlepage}
 \newgeometry{top=3cm,right=0cm,bottom=2cm,left=0cm}
\begin{center}
\fontsize{60pt}{72pt}\selectfont OptSCORE\par
\Large Group Communication in Dynamic Environments \par
\vskip 48pt
\Large Technical Report -- Milestone 1.1 \par
\Huge Analysis of Static and Dynamic Configurability \\ of Existing Group Communication Systems \par
\
\vfill
\Large \@date \par
\vskip 24pt
\Large \@author \par
\vskip 12pt
\Large Assistant Professorship of Security in Information Systems \\ University of Passau \par
\end{center}
\restoregeometry
\end{titlepage}
}
\makeatother

\begin{document}
\maketitle

\thispagestyle{empty}

\section*{Abstract}

Active replication following the state machine replication (SMR) approach is a way to make existing systems and services more reliable and fault-tolerant. The additional communication overhead has a negative impact on the system's throughput and overall request latency. Today's systems should be highly optimized to their execution environment and usage scenario in order to remedy the performance loss introduced by such group communication systems (GCS). In addition to that, systems should be able to adapt to changing environmental conditions. This report analyzes the available configuration options of three existing GCSs. Therefore, it explains the available configuration parameters and describes the given reconfiguration mechanisms. The found parameters are then classified in a parameter scheme. 

\newpage

\thispagestyle{plain}

\tableofcontents
\newpage

\input{src/introduction.tex}
\newpage
\input{src/gcs.tex}
\newpage
\input{src/parameters.tex}
\newpage
\input{src/conclusion.tex}


\newpage
\input{src/bibliography.tex}
\end{document}

%% file: src/introduction.tex
\section{Introduction} \label{Introduction}
This document presents the results of milestone 1.1 of the OptSCORE research project. The OptSCORE research project is a joint project between the University of Ulm and the University of Passau. In this milestone the static and dynamic configurability of existing Group Communication Systems (GCS) and State Machine Replication (SMR) systems have been analyzed \cite{optscore}.

\subsection{Motivation} \label{Motivation}
In recent years distributed services have become more and more complex. Similarly the non-functional requirements that they need to fulfill like security, reliability and performance have become more and more demanding. Services nowadays have to be replicated over the whole globe to provide low latencies to local clients and ensure high-availability. At the same time those services have to be implemented in a massive parallel manner to provide a high throughput and ensure scalability.

There are already several middleware solutions that apply the well-known SMR approach on top of GCS to support system architects and developers with fault-tolerance as well as reliability. Those systems add a significant communication overhead to the services that use them. Additionally, most systems require a sequential execution model to preserve the determinism of the underlying state machine, what goes hand-in-hand with a high request latency and a poor throughput performance. In order to remedy this issue, SMR systems are often highly optimized and combined with scheduling components that enable a deterministic multi-threaded execution.

Both components -- the group communication system as well as the scheduler -- entail a large set of configuration capabilities and their configurations must be heavily adapted to their execution environment in order to achieve an optimal runtime behavior. However, the average system developer might be overwhelmed by the sheer number of configuration parameters and most existing systems require a restart in order to carry out a reconfiguration. That's why many systems run with the default configuration that certainly might perform well in default situations, but misses out deviant situations and corner cases. The OptSCORE project tries to overcome those issues by inventing a GCS as well as a scheduling component which dynamically and autonomously adapt their configuration to the actual execution environment and hereby guarantee an optimal runtime behavior. The team at the University of Ulm is focused on the scheduler component, whereas the group communication system falls under the responsibility of the University of Passau.

\subsection{Objective} \label{Objective}
The goal of the first milestone included in the project's first work package (Dynamic Group Communication) is to analyze the static and dynamic configurability of existing group communication systems. Therefore we examine the general architecture as well as the configuration capabilities of existing group communication systems. During this examination configuration parameters as well as self-contained and substitutable building blocks are identified. These internal configuration capabilities are then augmented with configuration capabilities that control the execution environment and can only be adapted externally. A further analysis yields a detailed classification of parameters that on the one hand describe the state of a system as well as its environment at runtime and on the other hand parameters that can be used to configure the system components to influence the system execution. In this way the analysis will serve as a starting point for a more detailed examination taking place later in the project where for specific environmental conditions the optimal system configuration will be derived by executing series of experiments.


\subsection{Structure} \label{Structure}
The remainder of this document is structured as follows. The next section introduces existing group communications systems, investigates their architecture and analyzes their configuration parameters as well as their reconfiguration capabilities. After that, the following section introduces a parameter classification that categorizes the available parameters in monitoring and control parameters. These include both parameters internal to the application as well as parameters of the surrounding execution environment. Finally, the last section presents a short conclusion and gives an outlook for future work. 


%% file: src/gcs.tex
\section{Group Communication Systems} \label{Group Communication Systems}
This section analyzes existing GCSs and SMR systems. For each system the architecture is shortly summarized, before the configuration parameters that influence the performance of the system are listed. The description of these parameters is directly copied from the respective GCS documentation. That description is followed by a presentation of the reconfiguration capabilities and concluded with a short summary. One major characteristic of GCSs is their underlying fault model. We identified JGroups as representative for classical GCS and JPaxos as well as BFT-SMaRt as representatives for SMR systems with crash fault and Byzantine fault tolerance respectively.

\subsection{JGroups} \label{JGroups}
JGroups is a middleware framework written in Java that offers a highly customizable messaging service. It allows the creation and adaption of communication clusters by adding or removing computer nodes that are connected over a network. Nodes within a cluster form a view and its members are able to reliably pass unicast as well as multicast messages between each other and also to detect crashed nodes. For the failure detection JGroups assumes a crash-fault model \cite{jgroups}. At the time of writing JGroups was still under active development and the latest stable release was version 3.6.6 \cite{jgroups-code}.

\subsubsection{Architecture} \label{JGroups:Architecture}
The JGroups Application Programming Interface (API) provides different objects at different levels of abstractions. Protocols with their layered/stacked architecture form the basis of the communication framework. Starting from the lowest layer -- some kind of unreliable transport primitive -- every protocol layer adds some specific service like message ordering or group membership to the protocol, resulting in a feature-rich transportation protocol. Nodes use these protocols when communicating with other nodes. For this purpose there are two interfaces -- \lstinline{Channel} and \lstinline{RpcDispatcher}. A \lstinline{Channel} object can be used to send messages, whereas a \lstinline{RpcDispatchen} supports remote procedure calls. Besides that, there are some more sophisticated data structures built on top of channels. Representatives of the so called building blocks are distributed locks and counters as well as replicated caches or replicated hash maps.

\subsubsection{Configuration Parameters} \label{JGroups:Configuration Parameters}
Our analysis focuses on the protocols as they form the base components that all other components use and rely on. Thus the remainder of this section describes various protocol types and lists those configurations parameters that might influence the overall performance. The parameter descriptions are borrowed from the JGroups manual \cite{jgroups-manual}.

\begin{description}
	\item[Transport] The transport protocols define the transport type (i.e. UDP and TCP) and specify their execution properties. Besides other basic settings like the socket binding and the log setup it allows the fine adjustment of message handling (batching as well as bundling), buffer sizes, timeouts, and thread pool properties. 
	
	The TP protocol is used as a base for all high-order protocols.
	\begin{description}
		\item[enable\_batching] Allows the transport to pass received message batches up as MessagesBatch instances, rather than individual messages. This flag will be removed in a future version when batching has been implemented by all protocols
		\item[enable\_bundling] Enable bundling of smaller messages into bigger ones. Default is true
		\item[max\_bundle\_size] Maximum number of bytes for messages to be queued until they are sent
		\item[max\_bundle\_timeout] Max number of milliseconds until queued messages are sent
		\item[time\_service\_interval] Interval (in ms) at which the time service updates its timestamp. 0 disables the time service
		\item[internal\_thread\_pool\_enabled] Switch for enabling thread pool for internal messages
		\item[oob\_thread\_pool\_enabled] Switch for enabling thread pool for out-of-band (oob) messages. Default is true
		\item[internal/oob/timer\_thread\_pool\_min\_threads] Minimum thread pool size for the internal/oob/timer thread pool
		\item[internal/oob/timer\_thread\_pool\_max\_threads] Maximum thread pool size for the internal/oob/timer thread pool
	\end{description}
	
	UPD is the first real ready-to-use transport protocol. By default it uses IP multicast for multicast messages and UDP datagrams for unicast messages. Usually UDP is chosen as transport protocol if all cluster nodes reside in the same subnet, as IP multicast is most of the time not available across subnets.	
	\begin{description}
		\item[ip\_mcast] Multicast toggle. If false multiple unicast datagrams are sent instead of one multicast. Default is true
		\item[ip\_mcast] The time-to-live (TTL) for multicast datagram packets. Default is 8
		\item[mcast\_recv\_buf\_size] Receive buffer size of the multicast datagram socket. Default is 500'000 bytes
		\item[mcast\_send\_buf\_size] Send buffer size of the multicast datagram socket. Default is 100'000 bytes
		\item[ucast\_recv\_buf\_size] Receive buffer size of the unicast datagram socket. Default is 64'000 bytes
		\item[ucast\_send\_buf\_size] Send buffer size of the unicast datagram socket. Default is 100'000 bytes
		\item[tos] Traffic class for sending unicast and multicast datagrams. Default is 8
	\end{description}
	
	TCP/TCP\_NIO2 is the other type of transport protocol. Using that protocol every node maintains a connection to every other node from the cluster. Therefore sending a multicast message is in the complexity class of $O(n)$, whereas with UDP and IP multicast it is just $O(1)$. 
	
	\begin{description}
			\item[conn\_expire\_time] Max time connection can be idle before being reaped (in ms)
			\item[reaper\_interval] Reaper interval in msec. Default is 0 (no reaping)
			\item[recv\_buf\_size] Receiver buffer size in bytes
			\item[send\_buf\_size] Send buffer size in bytes
			\item[send\_queue\_size] Max number of messages in a send queue
			\item[sock\_conn\_timeout] Max time allowed for a socket creation in connection table
			\item[tcp\_nodelay] Should TCP no delay flag be turned on
			\item[use\_send\_queues] Should separate send queues be used for each connection
			\item[max\_read\_batch\_size] Only in TCP\_NIO2. Max number of messages a read will try to read from the socket. Setting this to a higher value will increase speed when receiving a lot of messages. However, when the receive message rate is small, then every read will create an array of max\_read\_batch\_size messages.
			\item[max\_send\_buffers] Only in TCP\_NIO2. The max number of outgoing messages that can get queued for a given peer connection (before dropping them). Most messages will be retransmitted; this is mainly used at startup, e.g. to prevent dropped discovery requests or responses (sent unreliably, without retransmission).
		\end{description}
		
	\item[Discovery] The discovery protocols are used to get current membership information from other cluster members. Hereby cluster members can discover each other, determine the identity of the current coordinator and detect diverging cluster membership information. Depending on the actual protocol the discovery procedure can take place in a static or a dynamic way, i.e. MPING uses IP multicast to send a Discovery request to every node in the subnet, whereas TCPPING relies on a predefined list of nodes that shall be contacted.  
	
	\begin{description}
		\item[num\_initial\_members] Minimum number of initial members to get a response from
		\item[timeout] Timeout to wait for the initial members
		\item[stagger\_timeout] If greater than 0, we’ll wait a random number of milliseconds in range [0..stagger\_timeout] before sending a discovery response. This prevents traffic spikes in large clusters when everyone sends their discovery response at the same time
	\end{description}
		
	\item[Merging] Different merge protocols are available to merge nodes back into a cluster after a network partition happened. The merging of the cluster's state is left to the actual application (JGroups provides a callback that has to be handled). MERGE2 is marked as deprecated. Instead MERGE3 should be used. In this protocol specific INFO messages are exchanged periodically to detect such partitions. Those messages contain the nodes' identity and its current view. Diverging views are used to detect cluster partitions that can be merged back together.
		
	\begin{description}
		\item[check\_interval] Interval (in ms) after which we check for view inconsistencies
		\item[max\_interval] Interval (in milliseconds) when the next info message will be sent. A random value is picked from range [1..max\_interval]
		\item[min\_interval] Minimum time in ms before sending an info message
	\end{description}
		
	\item[Failure Detection] The failure detection protocols allow the detection of node crashes. Therefore the protocols implement different strategies. The FD protocol aligns group members in a logical ring, where special heartbeat messages are sent along. 
	\begin{description}
		\item[max\_tries] Number of times to send an are-you-alive message
		\item[msg\_counts\_as\_heartbeat] Treat messages received from members as heartbeats. Note that this means we’re updating a value in a hashmap every time a message is passing up the stack through FD, which is costly.
		\item[timeout] Timeout to suspect a node P if neither a heartbeat nor data were received from P.
	\end{description}
	
	In the FD\_ALL and FD\_ALL2 protocol nodes multicast heartbeat nodes to all cluster nodes. Each node periodically checks the receipt of those messages in order to suspect nodes to be crashed.
	\begin{description}
		\item[interval] Interval at which a HEARTBEAT is sent to the cluster
		\item[timeout\_check\_interval] Timeout after which a node P is suspected if neither a heartbeat nor data were received from P
		\item[timeout\_check\_interval] Interval at which the HEARTBEAT timeouts are checked
	\end{description}
	
	The FD\_SOCK protocol follow a different approach. It does not send any heartbeat messages but uses the connection state of dedicated TCP sockets to detect crashes. Therefore the nodes form -- analogous to the FD\_ALL protocol -- a logical ring in which every nodes maintains one special TCP connection to its direct neighbors. If the connection is closed abruptly, the node at the other end of the communication link gets suspected. 
	\begin{description}
		\item[keep\_alive] Whether to use KEEP\_ALIVE on the ping socket or not. Default is true
		\item[suspect\_msg\_interval] Interval for broadcasting suspect messages. Default is 5000 ms
	\end{description}
	
	This protocol family also provides the VERIFY\_SUSPECT protocol that can be used as a last check before the unresponsive node is excluded from the current view. With that mechanism false suspicions can be reduced.
	\begin{description}
		\item[num\_msgs] Number of verify heartbeats sent to a suspected member
		\item[timeout] Number of millisecs to wait for a response from a suspected member
		\item[use\_mcast\_rsps] Send the I\_AM\_NOT\_DEAD message back as a multicast rather than as multiple unicasts (default is false)
	\end{description}

	\item[Reliable Transmission] The protocol family for the reliable message transmission ensures that no messages ever gets lost. The NAKACK protocol provides this property for multicasts by using sequential sequence numbers for all messages and negative acknowledgment messages. Such a message is sent whenever a gap in the sequence numbers occurs so that the lowest received message number does not directly follow the highest delivered number (i.e. if a process p delivered message q:1 and received message q:4 and q:3, it will send an NACK message asking for message 2 to process q).
	
	\begin{description}
		\item[become\_server\_queue\_size] Size of the queue to hold messages received after creating the channel, but before being connected (is\_server=false). After becoming the server, the messages in the queue are fed into up() and the queue is cleared. The motivation is to avoid retransmissions.
		\item[discard\_delivered\_msgs] Should messages delivered to application be discarded
		\item[max\_rebroadcast\_timeout] Timeout to rebroadcast messages. Default is 2000 ms
		\item[max\_xmit\_req\_size] Max number of messages to ask for in a retransmit request. 0 disables this and uses the max bundle size in the transport
		\item[resend\_last\_seqno] If enabled, multicasts the highest sent seqno every xmit\_interval ms. This is skipped if a regular message has been multicast, and the task aquiesces if the highest sent seqno hasn’t changed for resend\_last\_seqno\_max\_times times. Used to speed up retransmission of dropped last messages (JGRP-1904)
		\item[resend\_last\_seqno\_max\_times] Max number of times the last seqno is resent before acquiescing if last seqno isn’t incremented
		\item[use\_mcast\_xmit] Retransmit retransmit responses (messages) using multicast rather than unicast
		\item[use\_mcast\_xmit\_req] Use a multicast to request retransmission of missing messages
		\item[xmit\_interval] Interval (in milliseconds) at which missing messages (from all retransmit buffers) are retransmitted
	\end{description}
	
	The UNICAST protocol in contrast uses positive acknowledgment messages and does only support unicast messages. So if a process p sends a message m it has to cache that message m until it received an acknowledgment message from the receiver or information about its crash.
		
	\begin{description}
		\item[ack\_batches\_immediately] Send an ack for a batch immediately instead of using a delayed ack
		\item[ack\_threshold] Send an ack immediately when a batch of ack\_threshold (or more) messages is received. Otherwise send delayed acks. If 1, ack single messages (similar to UNICAST)
		\item[conn\_close\_timeout] Time (in ms) until a connection marked to be closed will get removed. 0 disables this
		\item[conn\_expiry\_timeout] Time (in milliseconds) after which an idle incoming or outgoing connection is closed. The connection will get re-established when used again. 0 disables connection reaping
		\item[max\_retransmit\_time] Max number of milliseconds we try to retransmit a message to any given member. After that, the connection is removed. Any new connection to that member will start with seqno \#1 again. 0 disables this
		\item[max\_xmit\_req\_size] Max number of messages to ask for in a retransmit request. 0 disables this and uses the max bundle size in the transport
		\item[xmit\_interval] Interval (in milliseconds) at which messages in the send windows are resent
	\end{description}
		
	\item[Message stability] The message stability protocols does the garbage collection for messages that have been cached at sender nodes to enable the retransmission of lost messages. Therefore all members exchange information about the highest sequence number of their messages received and hereby agree on messages that can be removed from their retransmission tables. This is done periodically or if a retransmission table uses a certain amount of space. 
		
	\begin{description}
		\item[desired\_avg\_gossip] Average time to send a STABLE message
		\item[max\_bytes] Maximum number of bytes received in all messages before sending a STABLE message is triggered
		\item[send\_stable\_msgs\_to\_coord\_only] Wether or not to send the STABLE messages to all members of the cluster, or to the current coordinator only. The latter reduces the number of STABLE messages, but also generates more work on the coordinator
		\item[stability\_delay] Delay before stability message is sent
	\end{description}
		
	\item[Group Membership] Group membership allows the formation and administration of group/clusters. This includes joining new clusters, leaving existing clusters and handling crashed cluster members.
		\begin{description}
			\item[join\_timeout] Join timeout
			\item[leave\_timeout] Leave timeout
			\item[merge\_timeout] Timeout (in ms) to complete merge
			\item[max\_join\_attempts] Number of join attempts before we give up and become a singleton. Zero means never give up.
			\item[view\_ack\_collection\_timeout] Time in ms to wait for all VIEW acks (0 means wait forever). Default is 2000 ms
		\end{description}
		
	\item[Flow Control] Messages that can be directly processed or cached at the receiver side will be dropped and need to be retransmitted which poses a costly procedure. The flow control protocols try to adjust the send rate of a sender to the receiver with the lowest reception capacity.
		\begin{description}
			\item[ignore\_synchronous\_response] Does not block a down message if it is a result of handling an up message in the same thread. Fixes JGRP-928
			\item[max\_block\_time] Max time (in milliseconds) to block. Default is 5000 ms
		\end{description}
		
	\item[Ordering] The ordered reception of messages might be crucial for certain applications. The SEQUENCER protocol uses a coordinator/sequencer to implement a total order multicast. There is also the experimental TOA protocol that uses the Skeen algorithm to provide total ordered anycast messages, so that a message sent to only a part of the cluster still fulfills the total order requirements.
		
	\item[State Transfer] State transfer protocols allow the synchronization of the nodes' state. Therefore a simple byte array is exchanged between the nodes as TCP stream or as multiple chunks. The configurability focusses on the state transfer buffer and on the worker threads handling the state transfers.
			\begin{description}
				\item[buffer\_size] Size (in bytes) of the state transfer buffer
				\item[max\_pool] Maximum number of pool threads serving state requests
			\end{description}
		
	\item[Fragmentation] The fragmentation protocols FRAG and FRAG2 simply fragment large messages into smaller chunks and reassemble them back together at the communication endpoint.
		\begin{description}
			\item[frag\_size] The max number of bytes in a message. Larger messages will be fragmented
		\end{description}
		
	\item[Compress] The COMPRESS protocol allows the compression and decompression before and after a message is sent between cluster nodes.
			\begin{description}
				\item[compression\_level] Compression level (from java.util.zip.Deflater) (0=no compression, 1=best speed, 9=best compression). Default is 9.
				\item[min\_size] Minimal payload size of a message (in bytes) for compression to kick in. Default is 500 bytes.
				\item[pool\_size] Number of inflaters/deflaters for concurrent processing. Default is 2.
			\end{description}
			
	\item[Security] There are several security protocols that provide security services to the cluster. Confidentiality is provided by the ENCRYPT protocol.
	
			\begin{description}
				\item[asymAlgorithm] Cipher engine transformation for asymmetric algorithm. Default is RSA
				\item[asymInit] Initial public/private key length. Default is 512
				\item[changeKeysOnViewChange] Generate new symmetric keys on every view change. Default is false. Set this to true when using asymmetric encryption, to handle merging (JGRP-1907)
				\item[symAlgorithm] Cipher engine transformation for symmetric algorithm. Default is AES
				\item[symInit] Initial key length for matching symmetric algorithm. Default is 128
			\end{description}
				
	Whereas authentication can be achieved with the use of the AUTH protocol. AUTH is an abstract protocol so its implementation is left to the actual application. However provides the SASL protocol as an alternative. With the SASL applications can use a SASL mechanism from the JDK.
		
	\item[Others] There are several other protocols that won't be discussed in detail, but only listed in this section for the sake of completeness.
		\begin{description}
			\item[BARRIER] The BARRIER protocol can be used to synchronize the execution.
			\item[FLUSH] The FLUSH protocol causes all cluster nodes to flush their pending request before a specific events happens. Hereby it can be ensured that the whole cluster processed a message. It is used for example in state transfers and view changes.
			\item[STATS] Some statistics can be collected using the STATS protocol. 
			\item[RELAY/RELAY2] The relay protocols allows bridging different cluster together.
			\item[RATE\_LIMITER] The RATE\_LIMITER -- as the name indicates -- allows restricting the data sending rate to a specific limit.
			\item[CENTRAL\_LOCK/PEER\_LOCK] Locking protocols can be used to implement a locking mechanism in the cluster. Locks can either be acquired and released at the cluster coordinator or by contacting all cluster nodes.
		\end{description}
\end{description}

\subsubsection{Reconfiguration Capabilities} \label{JGroups:Reconfiguration Capabilities}
The \lstinline{ProtocolStack} attribute of a \lstinline{Channel} object must be configured by either parsing an external eXtensible Markup Language (XML) file or by setting up a \lstinline{ProtocolStackConfigurator} object programmatically. This means on the one hand that new Channels can be introduced during the runtime. The new ones could be used as a replacement for the existing ones and hereby simulate a reconfiguration, but this needs some additional checkpointing. On the other hand, the protocol stack is not protected from modification, what means that every contained \lstinline{Protocol} can be retrieved and its properties can be changed at runtime. However the framework cannot make any guarantees what protocol reads the configuration properties at which time and if any unintended side effects take place. That's why we consider the direct reconfiguration possibilities at runtime just as a more theoretical possibility that also requires total knowledge about the whole protocol stack. 

\subsubsection{Summary} \label{JGroups:Summary}
To sum it up, the stacked architecture of JGroups provides a very flexible (and at the same time very complex) way to build tailor-made communication protocols. The performance of the resulting protocol is influenced by various similar configuration parameters. Grouping them together we see the following parameter classes: Worker threads and thread pool sizes, batching and bundling settings, compression and security services, buffer sizes and queue lengths, timeouts and repetitions, time intervals, number of group members or replicas, as well as the utilization of environment characteristics or the fine-tuning of protocol internals. These parameters will be described in more detail in Section \ref{Parameters}. Besides that we can conclude that JGroups generally provides extensive reconfiguration possibilities, even though we have to remark that those capabilities either require some synchronization checkpoints or some tentative integration deep into the application specific protocol stack.

\subsection{JPaxos} \label{JPaxos}
In contrast to the previous general purpose communication libraries, JPaxos is the first representative of Java-based GCSs that are solely dedicated to SMR. With the implementation of a MultiPaxos Protocol JPaxos is able to tolerate crash-recovery faults. Hereby, the implementation focuses on three main aspects: batching of decision requests and parallel consensus instances, view reconfiguration after the recovery of nodes, and multi-core-aware scaling \cite{jpaxos}. The latest stable release of JPaxos is version 1.0 and the project activity at the project's main source code repository is not indicative of active development \cite{jpaxos-code}.

\subsubsection{Architecture} \label{JPaxos:Architecture}
Due to its nature as SMR library JPaxos is based on three main components. The \lstinline{Service} object represents the application or service being replicated. Existing services are integrated into the middleware library by implementing several lifecycle callbacks that handle the execution of client request as well as the creation and utilization of application state snapshots. Besides that, there is the \lstinline{Replica} component that implements one replica of the service being replicated. And finally, there is the \lstinline{Client} component that hides the complexity of the communication -- like leader discovery and output consolidation -- whenever clients use the replicated service.
 
\begin{figure}[ht!]
	\centering	
		\includegraphics[width=1.0\linewidth]{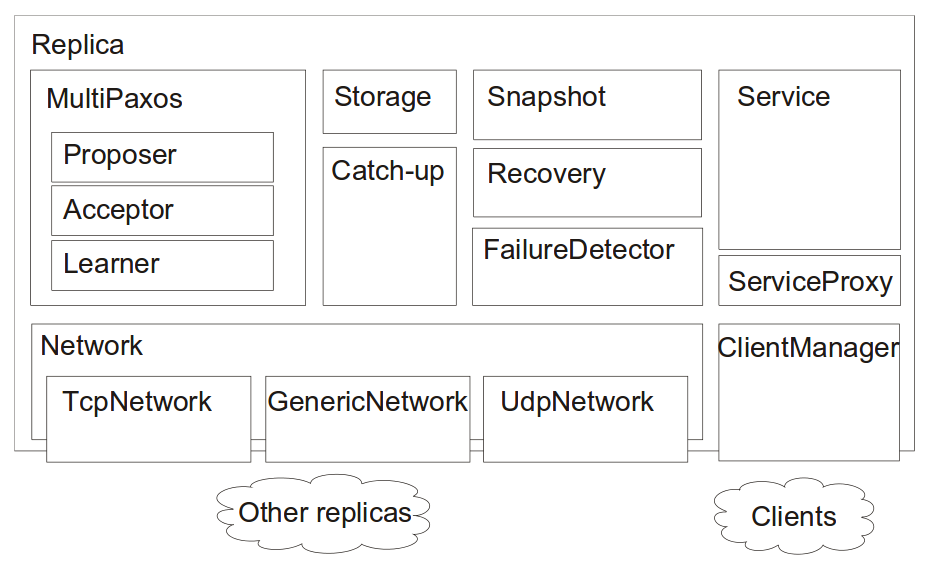}
	\caption{JPaxos Modules \cite{jpaxos}}
	\label{fig:JPaxos:Architecture}
\end{figure}

\subsubsection{Configuration Parameters} \label{JPaxos:Configuration Parameters}
JPaxos is configured using a plain-old Java properties files. During runtime this file is transformed in a \lstinline{Configuration} object, so that the particular system components can read its contents. The available configuration options are listed below.

\begin{description}
	\item[PROCESS] A node -- or a process in JPaxos terminology -- is configured with a single line containing hostname, replica port and client port, separated with commas:
	
	\begin{lstlisting}
	process.<id> = <hostname>:<replica_port>:<client_port>
	
	process.0 = localhost:2000:3000
	process.1 = localhost:2001:3001
	process.2 = localhost:2002:3002
	\end{lstlisting}
	
	Above configuration creates three replicas with ids: 0, 1, 2. Replica with id 0, is running on localhost and is using port 2000 to communicate with other replicas and is using port 3000 to accept connections from clients.
	
	\item[CRASH\_MODEL] One should select the crash model of the system. If the crash model uses the non-volatile memory (i.e. hard drive), the location of the logs may be specified as well.
	
	Currently supported crash models include:
	
	\begin{tabular}{llll}
	\textbf{Type}	& \textbf{Name}		& \textbf{Needs stable storage}	& \textbf{Fault tolerance} \\
	\midrule
	CrashStop		&	CrashStop			&	No						&	minority \\
	CrashRecovery	&	FullStableStorage	&	Yes, heavy usage		&	catastrophic \\
	CrashRecovery	&	ViewSS				&	Yes, periodically		&	minority \\
	CrashRecovery	&	EpochSS				&	Yes, one write by start	&	minority \\
	\end{tabular}
	
	Crash model types:
	\begin{description}
		\item[CrashStop] once the replica crashed, it cannot recover
		\item[CrashRecovery] the replica may crash and subsequently recover
	\end{description}
	
	Fault tolerance ranks:
	\begin{description}
		\item[minority] the minority of replicas may crash; $f=\lfloor(n-1)/2\rfloor$
		\item[catastrophic] all replicas may crash; $f=n$
	\end{description}
	
	To select a crash model, one must add to the configuration file a line CrashModel = [crash model name]. If no crash model is provided, the FullStableStorage is assumed.
	
	For choosing a log path one needs to add another line with syntax LogPath = [path]. The logs will be actually stored in subdirectory named after replica id in the given location. The default value for log path is jpaxosLogs.
	
	An example configuration:
	
	\begin{lstlisting}
	CrashModel = ViewSS
	LogPath = /mnt/shared/jpaxos/logs
	\end{lstlisting}
	
	\item[WINDOW\_SIZE] The window size determines the maximum number of concurrently proposed instances. The meaning of this option is very similar to window size in TCP protocol.
	
	To illustrate it, assume that window size is set to 10. It allows to run instances with id’s from 1 to 10 concurrently and to decide instances 2 - 10 before instance 1. JPaxos cannot execute any instance until all previous instances are executed so because instance 1 is not decided / executed, no instance can be executed on state machine. When instances 1 will be decided and executed all consecutive instances will also be executed.
	
	The example above shows that by increasing the value of window size we can decrease the response time - a lot of instances will be decided, but none can be executed. Because of that it is recommended to set this option to lower value and BatchSize to higher value so that decided instances can be executed faster.
	
	The default value of this option is 2.
	
	\item[BATCH\_SIZE] JPaxos will try to batch requests into a single proposal to improve the performance. This option controls the maximum size (in bytes) of requests grouped into one consensus instance.
	
	For example, if maximum batch size is set to 1000 and JPaxos received requests of size 100, 300, 400, 300 bytes, then first three requests will be batched into one consensus instance of size 100 + 300 + 400 = 800 (the size of all four requests is 1100 what is greater than maximum allowed batch size).

	The default value of this options is 65507 bytes.
	
	\item[MAX\_BATCH\_DELAY] This option determines how long JPaxos will wait for new requests to be packed into single instance. 
	
	The default value of this option is 10 ms.
	
	\item[MAX\_BATCH\_FETCHING\_TIME\_MS] How long can the proposer/catch-up wait for batch values during view change/catching up, in milliseconds. The default value is 2500 milliseconds.
	
	\item[INDIRECT\_CONSENSUS] If activated, the client request are batched and these batches are transported outside the actual Paxos protocol. This reduces the message size of leader messages sent in the Paxos Protocol. However, it does not reduce the overall network usage as the batches need to be transferred anyhow.
	
	By default this variant is not activated.
	
	
	\item[NETWORK] It is also possible to choose protocol used to communicate between replicas. One may choose:
	
	\begin{itemize}
		\item TCP (default)
		\item UDP
		\item Generic - Uses UDP for small messages and TCP for larger messages
	\end{itemize}

	It is important to note that UDP protocol has message size restriction - messages must be smaller than the maximum allowed size of UDP packet (64KB or less, depending on the network). User must be careful with the size of client requests and of batch size so that this limit is not violated. Because of this limitations, user should choose TCP or Generic option.
	
	If one chooses Generic, it is also recommended to set what is a ‘small’ and what is a ‘big’ message, by setting maximum allowed UDP packet size:
	
	\begin{lstlisting}
	Network = Generic
	MaxUDPPacketSize = 1000 
	\end{lstlisting}
	
	In example above, all messages smaller than 1000 bytes will be sent using UDP and all others using TCP protocol.
	
	\item[MAX\_UDP\_PACKET\_SIZE] Maximum UDP packet size in java is 65507. Higher than that and the send method throws an exception.
	
	In practice, most networks have a lower limit on the maximum packet size they can transmit. If this limit is exceeded, the lower layers will usually fragment the packet, but in some cases there's a limit over which large packets are simply dropped or raise an error.
	
	A safe value is the maximum Ethernet frame: 1500 - maximum Ethernet payload 20/40 - ipv4/6 header 8 - UDP header.
	
	Usually values up to 8KB are safe which is also the default value.
	
	\item[MTU] Specifies the maximum transmission unit (MTU) of the IP layer. The default size is 1492 bytes.

	\item[TCP\_RECONNECT\_TIMEOUT] Specifies how long it will be waited for performing a reconnect if a TCP connection fails. The default value is 1000 milliseconds.
			
	\item[CLIENT\_REQUEST\_BUFFER\_SIZE] Size of a buffer for reading client requests; larger requests than this size will cause extra memory allocation and freeing at each such request. This variable impacts memory usage, as each client connection pre-allocates such buffer. The default buffer size is 8212 kilobytes.
			
	\item[SELECTOR\_THREADS] How many selector threads to use. Selector threads are used to read and write requests from clients. Default value is -1.
	
	\item[FD\_SUSPECT\_TO] How long to wait until suspecting the leader. The default value is 1000 milliseconds
	
	\item[FD\_SEND\_TO] Interval between sending heartbeats. The default value is 1000 milliseconds.

	\item[SELECTOR\_THREADS] 
		
	\item[RETRANSMIT\_TIMEOUT] Specifies when an unhandled message will be retransmitted. The default retransmission timeout is 1000 milliseconds. 
		
	\item[SNAPSHOT\_MIN\_LOG\_SIZE] Minimum size of the log before a snapshot is attempted. The default value is 100 * 1024 kilobytes.
		
	\item[SNAPSHOT\_ASK\_RATIO] Ratio = $\frac{log}{snapshot}$. How bigger the log must be to ask. The default value is 1.
	
	\item[SNAPSHOT\_FORCE\_RATIO] Ratio = $\frac{log}{snapshot}$. How bigger the log must be to force. The default value is 2.
		
	\item[MIN\_SNAPSHOT\_SAMPLING] Minimum number of instances for checking ratios. The default value is 50.

	\item[FORWARD\_MAX\_BATCH\_SIZE] The maximum number of client-requests that are aggregated in one batch.
	
	 The default value is 1450.

	\item[FORWARD\_MAX\_BATCH\_DELAY] The maximum amount of time to wait for additional messages to include in a newly created batch.
	
	 The default value is 50 milliseconds.
	
\end{description}

\subsubsection{Reconfiguration Capabilities} \label{JPaxos:Reconfiguration Capabilities}
As already mentioned above the central configuration file is read during startup and transformed in a \lstinline{Configuration} object. This configuration is then consulted several times by different system components. Even if the configuration of those components can be adapted programmatically by modifying the source code, there are in general no interfaces to change the configuration options during runtime.

\subsubsection{Summary} \label{JPaxos:Summary}
The extracted parameters basically match those identified during analyzing JGroups. There are again parameters for thread pools and buffer sizes, batching settings, as well as timeouts and delays. Here those parameters are used in the very specific context of SMR to configure the protocol execution. However, there are no possibilities to change those configurations during runtime.

\subsection{BFT-SMaRt} \label{BFT-SMaRt}
BFT-SMaRt is another SMR library that equips arbitrary services with reliable communication and byzantine fault tolerance (BFT) \cite{bft-smart}. The Java framework is heavily influenced by the Practical Byzantine Fault Tolerance (PBFT) replication library \cite{pbft}. However, the authors claim several advantages like better performance and more efficient fault tolerance. The latest release of the library is version 1.0 beta, which is still under active development \cite{bftsmart-code}.

\subsubsection{Architecture} \label{BFT-SMaRt:Architecture}
\begin{figure}[ht!]
	\centering	
		\includegraphics[width=.75\linewidth]{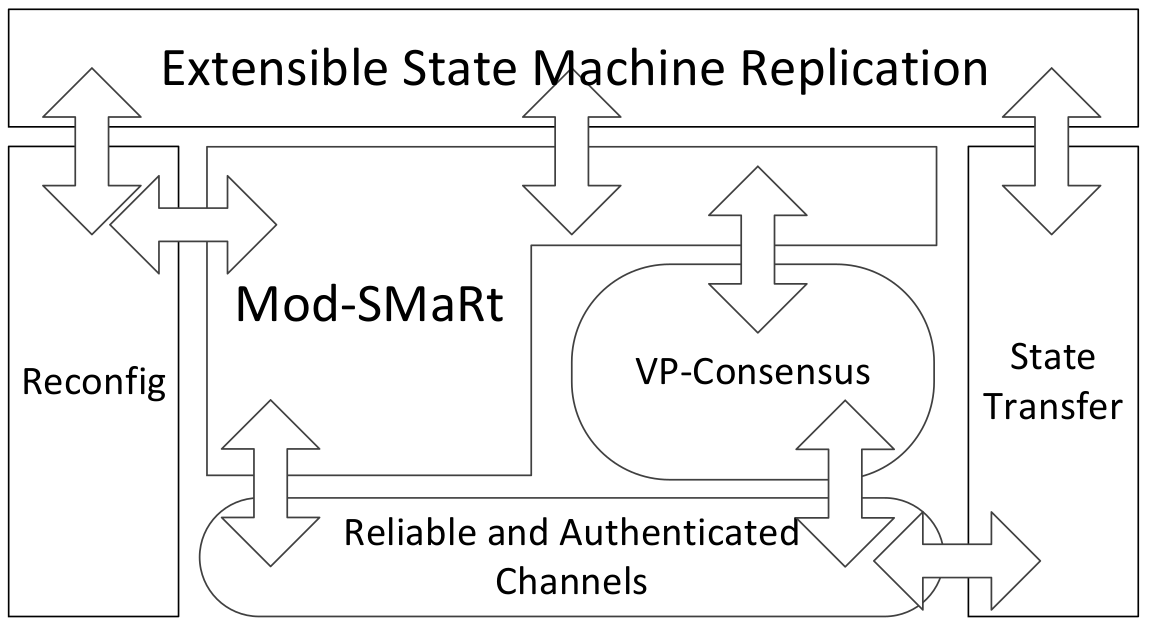}
	\caption{BFT-SMaRt Modules \cite{bft-smart}}
	\label{fig:BFT-SMaRt:Modules}
\end{figure}

The architecture of BFT-SMaRt follows a modular approach. As illustrated in Figure \ref{fig:BFT-SMaRt:Modules} the service being replicated is placed on top of the -- so called Mod-SMaRt -- SMR protocol. At the heart of this protocol operates a Validated and Proofable Consensus (VP-Consensus) implementation that is needed for determining the order of client request. In addition to that a reconfiguration component allows dynamic view changes during runtime and a state transfer protocol synchronizes the application state within the replica group.

\subsubsection{Configuration Parameters} \label{BFT-SMaRt:Configuration Parameters}
The configuration of the BFT-SMaRt library is separated in two configuration files. The first one defines the involved replicas. It serves as a lookup table for discovery purposes and lists all replicas together with their IP addresses and port numbers. The other file allows the configuration of the overall systems, such as the security settings or adjustments on the consensus procedure. The following list all available configuration parameters together with their official documentation and their default values \cite{bft-smart-config-documentation, bft-smart-config-example}.
\begin{description}
	\item[system.authentication.hmacAlgorithm] HMAC algorithm used to authenticate messages between processes (Default: HmacMD5)
	\item[system.communicatin.useSenderThread] Specify if the communication system should use a thread to send data (Default: true)
	\item[system.servers.num] Number of servers in the system (Default: 4)
	\item[system.servers.f] Maximum number of faulty replicas supported (Default: 1)
	\item[system.totalordermulticast.timeout] Time that a replica waits for the propose of a request. If the request is not proposed by the leader within the interval defined, the replica invokes the leader change protocol (Default: 10,000)
	\item[system.totalordermulticast.highMark] Maximum ahead-of-time message not discarded. When a replica is delayed it stores the consensus in an out-of-context queue. If the number of consensus to be processed is above the value defined in this parameter, the replicas discard the messages and invoke the state transfer protocol (Default: 10,000)
	\item[system.totalordermulticast.maxbatchsize] Maximum number of messages to be included in a single propose message (Default: 400)
	\item[system.total.ordermulticast.nonces] Number of nonces (for non-determinism actions) generated (Default: 0)
	\item[system.communication.inQueueSize] Number of messages that can be stored in the receive queue of the communication system (Default: 100,000)
	\item[system.communication.outQueueSize] Number of messages that can be stored in the send queue of the communication system (Default: 100,000)
	\item[system.communication.useSignatures] Used to define if clients should use signatures for MAC vectors (Default: 0)
	\item[system.communication.useMACs] Used for communication between the replicas. It defines if replicas should use authentication channels among them (Default: 0)
	\item[system.totalordermulticast.state\_transfer] Activates the state transfer protocol (Default: true)
	\item[system.totalordermulticast.revival\_highMark] Maximum ahead-of-time message not discarded when the replica is still on EID 0 (after which the state transfer is triggered). In practice this is used for a replica to verify if it crashed and was revived (Default: 10)
	\item[system.totalordermulticast.log] Stores the requests in a request log. (Default: true)
	\item[system.totalordermulticast.log\_parallel] Currently not used. (Default: false)
	\item[system.totalordermulticast.log\_to\_disk] Writes the log contents to a file on the local disk. (Default: false)
	\item[system.totalordermulticast.sync\_log] Writes every update to the log file synchronously to the underlying disk.
	\item[system.totalordermulticast.checkpoint\_period] Period at which the replica asks the state from the application and clear the log. Used to record a checkpoint and bound the size of the log. This number is the count of consensus messages processed (Default: 1000)
	\item[system.totalordermulticast.global\_checkpoint\_period] Period for the global checkpoint. (Default: 12000)
	\item[system.totalordermulticast.checkpoint\_to\_disk] Writes the checkpoint contents to a file on the local disk. (Default: false)
	\item[system.totalordermulticast.sync\_ckp] Writes every update to the checkpoint file synchronously to the underlying disk. (Default: false)
	\item[system.initial.view] Replicas IDs for the initial view, separated by comma. The number of replicas in this parameter should be equal to the one specified in system.servers.num (Default: 0,1,2,3)
	\item[system.ttp.id] The ID of the trusted third party (TTP). The TTP is used to add and remove replicas to the system (Default: 7002)
	\item[system.bft] This sets if the system will function in Byzantine or crash-only mode. Set to "true" to support Byzantine faults. (Default: true)
\end{description}

\subsubsection{Reconfiguration Capabilities} \label{BFT-SMaRt:Reconfiguration Capabilities}
The system configuration file is read during startup and then transformed into a \lstinline{TOMConfiguration} object. This object does not provide any means to change its values afterwards. Thus, there are no reconfiguration capabilities regarding the system configuration. The view, however, is configurable using a special client called ViewManager. This client is authorized to remove or add replicas to the system, without any further agreement decisions. 

\subsubsection{Summary} \label{BFT-SMaRt:Summary}
As with the JPaxos SMR library, most configuration options are related to the core SMR protocol. The particular parameters are similar to those already known from the other systems. Even if there are no possibilities to reconfigure the system configuration, there is the concept of a special client that is able to change the view during runtime.

%% file: src/parameters.tex
\section{Parameters} \label{Parameters}
This section describes the parameters that can be used to measure and configure the execution of a state machine replication system. First we define the parameters that can be measured to monitor the state of the system execution. After that, the configuration capabilities of such systems are listed. Therefore, the internal configuration parameters extracted during the analysis of existing GCSs in section \ref{Group Communication Systems} are augmented with common external configuration parameters that control the execution environment.


\subsection{Monitoring Parameters} \label{Monitoring Parameters}

Monitoring parameters measure the state of the system and its environment at runtime. They include host parameters, network parameters and application parameters. Host parameters describe the actual hardware utilization of the underlying computing resources. The network parameters describe the condition of the communication network that connects the nodes. And, application parameters characterize the application during its execution.

\subsubsection{Host Parameters} \label{Host Parameters}

\begin{description}
	\item[CPU Utilization] The central processing unit (CPU) utilization is a measure of the currently available processing resources. Usually a low CPU utilization indicates that the system is able to cope with the current workload. However, it can also be an indicator for suboptimal application of concurrency. It has to be noted that -- especially in shared environments -- the CPU is not dedicated solely to the monitored system.
	\item[RAM Utilization] Application data is stored in the random-access memory (RAM). The RAM utilization shows how much space is left in that memory. If the machine runs out of memory expensive paging operations may be executed. The RAM is analogous to the CPU not solely dedicated to the monitored system. Thus other processes running on the node might exhaust available memory.
	\item[Stable Storage Utilization] Data stored on stable storage outlasts power-off times and system restarts. For some applications it is crucial to protect data from those events by storing them on stable storage. If the available storage runs out, the system will become inconsistent or existing data will be overridden and must be recalculated.
	\item[Stable Storage Performance]
	The read and write performance can also become the system's bottleneck. If other tasks have to wait for write operations to be completed and those tasks take an unnecessary long time, the whole system's performance will go down. 
\end{description}

\subsubsection{Network Parameters} \label{Network Parameters}

\begin{description}
	\item[Data Rate] The data rate specifies the volume of data that can pass a medium during a specific time period. The higher the data rate, the more data can be send through the medium.  
	\item[Delay] The (nodal) delay gives information on the time a specific packet needs to travel from one node to another. The nodal delay is the sum of the following delays.
	\begin{description}
		\item[Processing Delay] Time that the node needs for processing the packet header and determine the next hop.
		\item[Queuing Delay] Time that the packet spends waiting in a queue to be transmitted. \\
		QueuingDelay = TransmissionDelay * QueueLength
		\item[Transmission Delay] Time needed to put a packet on the communication medium. \\
		TrasmissionDelay = PackageSize (bit) / TransmissionRate (bit/second)
		\item[Propagation Delay] Time that a signal needs to propagate inside the communication medium from one to the next node. \\
		PropagationDelay = DistanceFromNodeToNode (m) / PropagationSpeed (m/s)
	\end{description}
	The nodal delay is influenced by many (uninfluenceable) factors, including the geographical distance between the communication endpoints, temporary router congestions or changes in the routing of the packets.
	\item[Delay variation] Packet delay variation -- also called Jitter -- gives information about the latency variance of successive packet transmissions. A high variance makes it harder to adapt the packet transmission to some strategy as the delay is hard to predict.
	\item[Packet Loss Rate] Ratio between the lost packets and the packets sent in total. Lost packages cause high costs as they need to be retransmitted after their loss has been detected.
	\item[Packet Corruption Rate] Ratio between corrupted packets and the total number of packets. Corrupted packets cause high cost as they need to be retransmitted. Moreover the corruption might be noticed only at a higher application level.
	\item[Packet Rearrangement Rate] Ratio between the packets that arrive out-of-order and the total number of packets sent. Packet arriving out-of-order might be costly as they cause some reordering overhead and delay the delivery of packets sent earlier.
	\item[Packet Duplication Rate] Ration between duplicated packets and packets sent in total. Duplicated Packets may cause a moderate overhead as they need to be detected somewhere in the network stack.
	\item[Disconnects] Connection-oriented network protocols like TCP provide reliable packet transmission by relying on the establishment of a connection between both communication ends using a handshake protocol. Those connections are reused for subsequent packet transmission, but external factors may cause connection breaks -- or disconnects -- which negatively influences the packet transmission as the handshake between both communication ends has to be performed again.
	\item[Handoff/Handover] If mobile clients in mobile communication networks leave the range of its current cell and enter the area of another cell, the connection to the communication network needs to be transferred from the one to the other cell. This can cause additional delays or even the temporary loss of connectivity. 
\end{description}

\subsubsection{Application Parameters} \label{Application Parameters}

\begin{description}
	\item[Request Characteristics] Request can have various different characteristics and requirements that are listed in the following:
	\begin{description}
		\item[Size] The size of a request obviously influences the request's latency. The effect of the request size rises and falls with the data rate of the underlying communication network. 
		\item[Type] One important property of requests in the context of distributed state machines indicates if the request actually changes the internal state of the state machine -- if it is a write or a read-only request. Read-only requests can be executed out-of-order on different replicas, whereas the order of the write request has to be synchronized over the various replicas and each write operation must guarantee atomic consistency (linearizability).
		\item[Complexity] An application will need more time for processing complex requests as for simpler requests. This application specific characteristic has also an impact on the communication system, as queue sizes might need to be increased or requirements on the transmission delay might be lowered.
		\item[Frequency of Requests] Every request processor can be cluttered with a superior number of request, so that it has to quit its service -- even if the requests are small and simple. Insofar the frequency of requests is crucial if the request handling of an application or a service shall be analyzed.  
	\end{description}
	\item[Faults] In general faults take place whenever the system does not correspond to the specification. However for fault-tolerant systems it is important to define the fault model that the system is able to withstand. We generally distinguish between benign and malicious faults. Benign faults cause the affected node to crash and instantaneously quit its service. Thus a node either provides its service or not. While nodes that are affected by malicious or Byzantine faults can still provide their service but behave arbitrary. Crash faults can be further distinguished by their recovery model. In the crash-stop model nodes may fail by crashing and do not recover. Those nodes are removed from the group and won't be added back at some later point in time. Whereas in the crash-recovery model, nodes are added back to the group after they recovered \cite{groupcommunication,reliabledistributedprogramming}. Depending on the fault model faults can have the following characteristics.
	\begin{description}
		\item[Frequency] The frequency of faults defines their occurrences over time. For one single node a high fault frequency must not necessarily be worse than a moderate frequency as some faulty nodes as a whole might be tolerable and won't prevent the overall system from making progress.
		\item[Recovery Time] If the fault model supports the recovery of nodes, the recovery time determines the average duration of the time period the node needs to recover.
	\end{description}
	\item[System Utilization] The utilization of internal system resources can also give detailed insights about the health state of the system. However those parameters are highly application-specific.
	\begin{description}
		\item[Thread Pools] The variation of thread pool sizes over time may be a good indicator if the system can cope with the confronted system load and if the work load can be parallelized.
		\item[Caches] Caches include all types of queues and buffers and are usually used to cache work packages that cannot be executed directly or to cache objects that shall be grouped together. Analogous to thread pools, the system utilization can also be deducted from variations in the structure of the caches over time.
	\end{description}
\end{description}

\subsubsection{Optimization Parameters} \label{Optimization Parameters}
There are two further application parameters that are listed separately due to their special relevance in the project. These parameters can be used to measure the performance of the system -- and therefore the effectiveness of applied optimization strategies.

\begin{description}
	\item[Latency] Latency in communication networks is the time that a request takes from its source to its destination. The source and destination node must not necessarily be different nodes. Thus, in OptSCORE our primary goal is to reduce the latency of client request, which of course includes the reduction of latencies that are somehow related to them. 
	\item[Throughput] The other optimization goal is the throughput of client request. Various optimization strategies shall enable a significant increase of the throughput. This criteria can be measured in the number of requests or in the amount of processed bytes as well. 
\end{description}

\subsection{Control Parameters} \label{Control Parameters}
Control parameters configure specific system components to influence the system execution. The following section presents the internal and external control parameters that have been extracted from the analysis of existing group communication systems. They can be set initially or even changed during the runtime to control and optimize the runtime behavior of the GCS.

\subsubsection{Internal Control Parameters} \label{Internal Control Parameters}

\begin{description}
	\item[Worker] Various protocols allow the definition of thread pools sizes or the number of worker threads. The obvious intention is that with more executors more protocol steps can be executed (simultaneously). However it has to be noted that more executors produce additional administration overhead and that parallel execution will require synchronization efforts and may introduce deadlocks.
	\item[Batching/Bundling] Batching/Bundling is a common optimization strategy in distributed system. Instead of transferring messages on their own, messages are bundled together and sent. Thus bootstrapping efforts that normally arise for every single message are then only needed for the whole bundle. In state machine replication systems the order of client requests are usually decided in batches.
	\item[Compression] Data compression aims at reducing the size of the data. Compressed data is usually transferred faster over the communication medium. However the compression and decompression operations also take some time, so that there is some breakeven point for applying compression.
	\item[Intervals] In complex systems there are many long-lasting tasks -- mostly housekeeping -- that are triggered automatically at regular intervals. Those include garbage collection, buffer cleaning, or pool size consolidation. The exact intervals must be chosen to ensure both a healthy system and keeping the system load at a moderate level.
	\item[Timeouts] Timeouts are needed whenever the system might not behave as specified to ensure the liveness of the system. For instance a connection timeout ensures that the system does not wait for a response from the other communication end forever, if that communication end does not respond. Timeout are used in all variations: retransmission timeouts, batching timeouts, connection timeouts, socket creation timeout, discovery timeouts or heartbeat timeout.
	\item[Repetitions] Repetitions specify the tenacity of the system. It specifies how often message are retransmitted or how often components are probed before the are considered as failed. High repeptition rates may tolerate short periods of instability but may increase the overall latency of the system. 
	\item[Caches] Buffer sizes and queue lengths determine how many data can be temporarily stored. Large caches may improve the system performance as data items can be stored temporarily and do not need to be retransmitted.
	\item[Environment Exploitation] Some system make usage of specific environments. For instance some protocols can be configured to use IP multicast directly if the administrator can ensure that all group members are connected via some network that supports that feature.
	\item[Members/Replicas] The number of nodes that take part in the communication has various implications to the system's behavior. A higher number of nodes causes higher communication efforts, but it can also allow higher tolerance against faults and crashes. In some applications more members might also result in a lower client request latency, if the nodes are located geographically closer to the client.
	\item[Security] Security requirements are normally dictated from some external instance. However, the right selection of procedures and algorithms has a great impact on the overall system. 
	\item[Substitutable Components] Complex systems are composed of various self-contained components. For some components exist competitive implementations that can be selected through the configuration. For instance the transport layer may function either with UDP as well as with TCP communication, or for a total order multicast there might be different Consensus algorithms to choose from.
\end{description}

\subsubsection{External Control Parameters} \label{External Control Parameters}
\begin{description}
	\item[CPU Frequency] The CPU frequency defines the clock rate of the CPU, which in turn indicates how many arithmetic operations can be executed during a specific period. The higher the frequency the more (high-ordered) operations can be executed.
	\item[CPU Cores] In the simplest case a computer program is executed sequentially on one CPU core. If the program utilizes threads, it can be executed on multiple CPU cores in parallel. However doubling the number of CPU cores does not necessarily mean that the execution time will be halved. The program itself and its current execution has to support the utilization of every available thread equally.
	\item[RAM Size] The data structures needed during the program execution are stored in the volatile memory. The bigger the available RAM size, the more objects and data structures can be stored in the memory before the machine runs out of memory or probably long-lasting paging operations take place.
	\item[Stable Storage Size] Some application store crucial data on the stable storage. The size of that storage limits the data that can be stored.
	\item[Stable Storage Performance] Different types of stable storages have very different read and write performances. For instance the read and write performance of Solid State Drives (SSD) notably outperforms the performance of conventional Hard Disk Drives (HDD).
\end{description}

%% file: src/conclusion.tex
\section{Conclusion} \label{Conclusion}
This document presented the results of the first milestone. We analyzed existing GCSs and presented a complete list of their parameters. We classified those parameters under two main categories. Parameters that can be used to monitor the current state of the system execution and parameters that can be used to control the system at runtime. These results form the basis for future work, in which we are going to execute an exhaustive test series in order to see which parameters influence each other and to define optimal configurations for specific use cases and environmental conditions.


%% file: src/bibliography.tex
\bibliographystyle{splncs}
\bibliography{src/literature}

%% file: ms.bbl
\begin{thebibliography}{10}

\bibitem{optscore}
{Hauck, Franz J. and Reiser, Hans P.}:
\newblock {OptSCORE: Optimierung und Selbstkonfiguration von Kommunikation und
  Scheduling für replizierte Dienste}.
\newblock [online],
  \url{http://www.fim.uni-passau.de/sis/forschung/forschungsprojekte/?tx_importconveris_pi1[action]=showSingleProject&tx_importconveris_pi1[id]=1126},
  (Accessed: 10/11/15) (2014)

\bibitem{jgroups}
{Red Hat, Inc.}:
\newblock {JGroups}.
\newblock [online], \url{http://www.jgroups.org/}, (Accessed: 10/11/15) (2015)

\bibitem{jgroups-code}
{Github Inc.}:
\newblock {JPaxos / JPaxos}.
\newblock [online], \url{https://github.com/belaban/JGroups}, (Accessed:
  22/11/15) (2015)

\bibitem{jgroups-manual}
{Bela Ban}:
\newblock {Reliable group communication with JGroups}.
\newblock [online], \url{http://jgroups.org/manual/}, (Accessed: 10/11/15)
  (2015)

\bibitem{jpaxos}
Kończak, J., de~Sousa~Santos, N.F., Żurkowski, T., Wojciechowski, P.T.,
  Schiper, A.:
\newblock {JP}axos: {S}tate machine replication based on the {P}axos protocol.
\newblock Technical report (2011)

\bibitem{jpaxos-code}
{Github Inc.}:
\newblock {JPaxos / JPaxos}.
\newblock [online], \url{https://github.com/JPaxos/JPaxos}, (Accessed:
  22/11/15) (2015)

\bibitem{bft-smart}
{Bessani, A. and Sousa, J. and Alchieri, E.E.P.}:
\newblock {State Machine Replication for the Masses with BFT-SMART}.
\newblock In: Dependable Systems and Networks (DSN), 2014 44th Annual IEEE/IFIP
  International Conference on. (June 2014)  355--362

\bibitem{pbft}
{Castro, Miguel and Liskov, Barbara}:
\newblock {Practical Byzantine Fault Tolerance}.
\newblock In: Proceedings of the Third Symposium on Operating Systems Design
  and Implementation. OSDI '99, Berkeley, CA, USA, USENIX Association (1999)
  173--186

\bibitem{bftsmart-code}
{Github Inc.}:
\newblock {BFT-SMaRt}.
\newblock [online], \url{https://github.com/bft-smart/library}, (Accessed:
  22/11/15) (2015)

\bibitem{bft-smart-config-documentation}
{Github Inc.}:
\newblock {BFT SMaRt Configuration}.
\newblock [online],
  \url{https://github.com/bft-smart/library/wiki/BFT-SMaRt-Configuration},
  (Accessed: 20/11/15) (2015)

\bibitem{bft-smart-config-example}
{Github Inc.}:
\newblock {library/system.config at master · bft-smart/library}.
\newblock [online],
  \url{https://github.com/bft-smart/library/blob/master/config/system.config},
  (Accessed: 20/11/15) (2015)

\bibitem{groupcommunication}
Schiper, A.:
\newblock Group communication: From practice to theory.
\newblock In: {SOFSEM} 2006: Theory and Practice of Computer Science, 32nd
  Conference on Current Trends in Theory and Practice of Computer Science,
  Mer{\'{\i}}n, Czech Republic, January 21-27, 2006, Proceedings. (2006)
  117--136

\bibitem{reliabledistributedprogramming}
Guerraoui, R., Rodrigues, L.:
\newblock Introduction to reliable distributed programming (2006)

\end{thebibliography}
